\documentclass[%
	conference,
]{IEEEtran}

\usepackage[%
IEEE,%
]{copyright-overlay}

\CopyrightAuthor{Robert Falkenberg}
\CopyrightYear{2018}
\ConferenceName{2018 IEEE 88th Vehicular Technology Conference (VTC-Fall)}
\Doi{10.1109/VTCFall.2018.8690629}
\Bibtex{%
@InProceedings\{Falkenberg/etal/2018b,\textCR\textLF
\textHT{}author         = \{Robert Falkenberg and Benjamin Sliwa and Nico Piatkowski and Christian Wietfeld\},\textCR\textLF
\textHT{}title          = \{Machine Learning Based Uplink Transmission Power Prediction for \{LTE\} and Upcoming \{5G\} Networks Using Passive Downlink Indicators\},\textCR\textLF
\textHT{}booktitle      = \{2018 IEEE 88th Vehicular Technology Conference (VTC-Fall)\},\textCR\textLF
\textHT{}year           = \{2018\},\textCR\textLF
\textHT{}pages          = \{1-7\},\textCR\textLF
\textHT{}address        = \{Chicago, USA\},\textCR\textLF
\textHT{}month          = \{Aug\},\textCR\textLF
\textHT{}doi            = \{10.1109/VTCFall.2018.8690629\},\textCR\textLF
\textHT{}issn           = \{1090-3038\}\textCR\textLF
\}	
}

\newcommand{\datapath}{fig}

\usepackage{graphicx}
\usepackage{epstopdf}
\usepackage{import}
\makeatletter
\def\relativepath{\import@path}
\makeatother

\usepackage{standalone}
\usepackage[nolist ]{acronym}
\usepackage{tcolorbox}

\usepackage{xstring}

\usepackage[utf8]{inputenc}
\usepackage{marvosym}

\usepackage{sansmath}

\usepackage{algorithm}
\usepackage{algpseudocode}
\usepackage{tikz}

\usetikzlibrary{decorations.pathreplacing,decorations.markings,positioning,shapes,calc,shapes.arrows,patterns,backgrounds,fit,arrows,arrows.meta, shadows}
\tikzstyle{background grid}=[draw, black!50,step=1cm]	%
\usepackage{listings}
\definecolor{ListingBackground}{rgb}{0.97,0.97,0.97}
\usepackage{pgfplots}
\pgfplotsset{compat=newest}
\pgfplotsset{
    box plot/.style={
        /pgfplots/.cd,
        fill=blue!30,
        only marks,
        mark=-,
        mark size=0.2em,
        /pgfplots/error bars/.cd,
        y dir=plus,
        y explicit,
    },
    box plot box/.style={
        /pgfplots/error bars/draw error bar/.code 2 args={%
            \draw  ##1 -- ++(.2em,0pt) |- ##2 -- ++(-.2em,0pt) |- ##1 -- cycle;
        },
        /pgfplots/table/.cd,
        y index=2,
        y error expr={\thisrowno{3}-\thisrowno{2}},
        /pgfplots/box plot
    },
    box plot top whisker/.style={
        /pgfplots/error bars/draw error bar/.code 2 args={%
            \pgfkeysgetvalue{/pgfplots/error bars/error mark}%
            {\pgfplotserrorbarsmark}%
            \pgfkeysgetvalue{/pgfplots/error bars/error mark options}%
            {\pgfplotserrorbarsmarkopts}%
            \path ##1 -- ##2;
        },
        /pgfplots/table/.cd,
        y index=4,
        y error expr={\thisrowno{2}-\thisrowno{4}},
        /pgfplots/box plot
    },
    box plot bottom whisker/.style={
        /pgfplots/error bars/draw error bar/.code 2 args={%
            \pgfkeysgetvalue{/pgfplots/error bars/error mark}%
            {\pgfplotserrorbarsmark}%
            \pgfkeysgetvalue{/pgfplots/error bars/error mark options}%
            {\pgfplotserrorbarsmarkopts}%
            \path ##1 -- ##2;
        },
        /pgfplots/table/.cd,
        y index=5,
        y error expr={\thisrowno{3}-\thisrowno{5}},
        /pgfplots/box plot
    },
    box plot median/.style={
        /pgfplots/box plot
    },
    boxplot/every median/.style={
    	ultra thick,dashed,cyan
    }
}

\definecolor{flexicolor}{RGB}{46,49,146}
\definecolor{amaricolor}{RGB}{237,28,36}

\usepackage{xspace}

\usepackage[binary-units=true]{siunitx}
\usepackage{psfrag}
\usepackage{graphicx}
\usepackage{tabularx,booktabs}
\usepackage{multirow}
\usepackage{rotating}

\renewcommand{\baselinestretch}{1}

\usepackage{multirow}

\usepackage[cmex10]{amsmath}
\usepackage{amssymb}
\usepackage[caption=false,font=footnotesize]{subfig}

\usepackage{stfloats}

\usepackage{nicefrac}

\usepackage[]{acronym}
\usepackage{xspace}

\newcommand{\modF}{\ensuremath{\mathbb{F}}\xspace}
\newcommand{\modPI}{\ensuremath{\mathbb{P1}}\xspace}
\newcommand{\modPII}{\ensuremath{\mathbb{P2}}\xspace}
\newcommand{\modS}{\ensuremath{\mathbb{S}}\xspace}

\newcommand{\DeepL}{Deep Learning\xspace}
\newcommand{\RandomF}{Random Forest\xspace}
\newcommand{\RidgeR}{Ridge Regression\xspace}

\newcommand{\ML}{machine learning\xspace}

\acrodef{CI}{confidence interval}
\newcommand{\CI}{\ac{CI}\xspace}
\newcommand{\CIs}{\acp{CI}\xspace}

\acrodef{TX-power}{transmission power}
\newcommand{\txpower}{\ac{TX-power}\xspace}

\acrodef{HTTP}{Hypertext Transfer Protocol}
\newcommand{\HTTP}{\ac{HTTP}\xspace}

\acrodef{V2X}{Vehicle-to-everything}

\acrodef{TPC}{Transmission Power Control}
\newcommand{\TPC}{\ac{TPC}\xspace}

\acrodef{SVM}{Support Vector Machine}

\acrodef{DT}{Decision Tree}

\acrodef{RBF}{radial basis function}

\acrodef{RFC}{Random Forest Classifier}

\acrodef{KNN}{K-Nearest-Neighbour}

\acrodef{MAP}{Maximum a Posterior}

\acrodef{GNB}{Gaussian Naive Bayes}

\acrodef{EQRR}{Effective Query Response Ratio}

\acrodef{QRR}{Query Response Ratio}

\acrodef{EQRT}{Effective Query Response Time}

\acrodef{FRAM}{Ferroelectric Random Access Memory}

\acrodef{ETSI}{European Telecommunications Standards Institute}

\acrodef{QRT}{Query Response Time}

\acrodef{PSDR}{Product Specific Delivery Ratio}

\acrodef{PTA}[PTA]{Priced Timed Automata}

\acrodef{CSMACA}[CSMA/CA]{Carrier Sense Multiple Access / Collision Avoidance}

\acrodef{CCA}[CCA]{Clear Channel Assessment}

\acrodef{CSV}{comma separated values}

\acrodef{GFSK}[GFSK]{Gaussian frequency-shift keying}

\acused{GFSK}	%

\acrodef{GPS}[GPS]{Global Positioning System}
\acrodefplural{GPS}[GPS]{Global Positioning Systems}

\acrodef{MAC}{Media Access Control}

\acrodef{MCU}{Micro Controller Unit}

\acrodef{CPS}[CPS]{Cyber Physical System}
\acrodefplural{CPS}[CPS]{Cyber Physical Systems}

\acrodef{WSN}[WSN]{Wireless Sensor Networks}

\acrodef{AP}[AP]{Access Point}

\acrodef{BLE}{Bluetooth Low Energy}

\acrodef{SSB}[SSB]{Swappable Slave Board}

\acrodef{OMNeT++}[OMNeT++]{Objective Modular Network Testbed in C++}

\acrodef{LBT}[LBT]{Listen Before Talk}

\acrodef{SRD}[SRD]{Short Range Devices}

\acrodef{LTE}[LTE]{Long Term Evolution}
\newcommand{\LTE}{\ac{LTE}\xspace}

\acrodef{LTE-A}[LTE-A]{LTE-Advanced}
\newcommand{\LTEA}{\ac{LTE-A}\xspace}

\acrodef{LQI}{Link Quality Indicator}

\acrodef{MEMS}[MEMS]{Microelectromechanical system}

\acrodef{ECDF}[ECDF]{Empirical Cumulative Distribution Function}

\acrodef{PDU}{packet data unit}

\acrodef{RFID}{Radio-Frequency Identification}

\acrodef{SPI}{Serial Peripheral Interface}

\acrodef{TDMA}{Time-Division Multiple Access}

\acrodef{UWB}{Ultra-wideband}

\acrodef{RSSI}{Received Signal Strength Indicator}
\acrodef{RSRP}{Reference Signal Received Power}
\acrodef{RSRQ}{Reference Signal Received Quality}
\acrodef{SNR}{Signal to Noise Ratio}
\acrodef{SINR}{Signal to Interference and Noise Ratio}
\acrodef{MCS}[MCS]{Modulation and Coding Scheme}

\acrodef{TBS}[TBS]{Transport Block Size}

\acrodef{PRB}[\text{PRB}]{Physical Resource Block}
\acrodef{RB}[\text{RB}]{Resource Block}
\acrodef{CRC}[CRC]{Cyclic Redundancy Check}
\acrodef{DCI}[DCI]{Downlink Control Information}
\acrodef{RNTI}[RNTI]{Radio Network Temporary Identifier}
\acrodef{SI-RNTI}[\ensuremath{\mathrm{SI-RNTI}}]{System-Information \acs{RNTI}}
\acrodef{CCE}[CCE]{Control Channel Element}
\acrodef{PDCCH}[PDCCH]{Physical Downlink Control Channel}
\acrodef{UE}[UE]{User Equipment}
\acrodef{eNodeB}[eNodeB]{evolved NodeB}
\acrodef{STG}[STG]{Smart Traffic Generator}
\acrodef{DUT}[DUT]{Device Under Test}
\acrodef{OAI}[OAI]{Open Air Interface}
\acrodef{OFDM}[OFDM]{Orthogonal Frequency Division Multiplexing}
\acrodef{DL}[DL]{Downlink}
\acrodef{UL}[UL]{Uplink}
\acrodef{C3ACE}[C\textsuperscript{3}\!ACE\xspace]{Client-based Control Channel Analysis for Connectivity Estimation}
\acrodef{EC3ACE}[E-C\textsuperscript{3}\!ACE\xspace]{Enhanced Client-based Control Channel Analysis for Connectivity Estimation}
\acrodef{TCP}[TCP]{Transmission Control Protocol}
\newcommand{\TCP}{\ac{TCP}\xspace}
\acrodef{UDP}[UDP]{User Datagram Protocol}
\newcommand{\UDP}{\ac{UDP}\xspace}
\acrodef{FTP}[FTP]{File Transfer Protocol}

\acrodef{CoPoMo}{Context-aware Power Consumption Model}
\newcommand{\copomo}{\ac{CoPoMo}\xspace}
\acrodef{BSE}{Base Station Emulator}

\acrodef{CA}{Carrier Aggregation}

\acrodef{CC}{Component Carrier}

\acrodef{MIMO}{Multiple Input Multiple Output}

\acrodef{SISO}{Single Input Single Output}

\acrodef{AWGN}{Additional White Gaussian Noise}

\acrodef{COTS}{Commercial Off-the-Shelf}
\newcommand{\COTS}{\ac{COTS}\xspace}
\acrodef{SDR}{Software-Defined Radio}
\newcommand{\SDR}{\ac{SDR}\xspace}
\acrodef{IoT}{Internet of Things}
\newcommand{\IoT}{\ac{IoT}\xspace}
\acrodef{SDK}{Software Development Kit}
\acrodef{UBP}{Unit Backoff Period}

\acrodef{PCC}[PCC]{Primary Carrier Component}

\acrodef{SCC}[SCC]{Secondary Carrier Component}

\newcommand{\ECDF}{\ac{ECDF}\xspace}
\newcommand{\RSSI}{\ac{RSSI}\xspace}
\newcommand{\RSRP}{\ac{RSRP}\xspace}
\newcommand{\RSRQ}{\ac{RSRQ}\xspace}

\newcommand{\SINR}{\ac{SINR}\xspace}
\newcommand{\PRB}{\ac{PRB}\xspace}

\newcommand{\RB}{\ac{RB}\xspace}
\newcommand{\RBs}{\acp{RB}\xspace}

\newcommand{\UE}{\ac{UE}\xspace}
\newcommand{\UEs}{\acp{UE}\xspace}
\newcommand{\eNB}{\ac{eNodeB}\xspace}

\newcommand{\DL}{\ac{DL}\xspace}
\newcommand{\UL}{\ac{UL}\xspace}

\acrodef{RMSE}{Root Mean Square Error}
\newcommand{\RMSE}{\ac{RMSE}\xspace}
\acrodef{MAE}{Mean Absolute Error}
\newcommand{\MAE}{\ac{MAE}\xspace}

\acrodef{API}{Application Programming Interface}
\newcommand{\API}{\ac{API}\xspace}
\acrodef{APIs}{Application Programming Interfaces}

\newcommand{\fullModel}{full-featured model\xspace}
\newcommand{\practicalModel}{practical model\xspace}
\newcommand{\simulationModel}{simulation model\xspace}

\acresetall

\newcommand{\paperTitle}{Machine Learning Based Uplink Transmission Power Prediction for LTE and Upcoming 5G Networks using Passive Downlink Indicators}

\newcommand{\figurePadding}{0pt}
\newcommand{\figureTopPadding}{\figurePadding}
\newcommand{\figureBottomPadding}{-15pt}

\newcommand{\cf}{cf.\xspace}		%
\newcommand{\eg}{e.g.\xspace}		%

\newcommand{\dB}{\decibel}
\newcommand{\dBm}{dBm}

\newcommand{\Fig}[1]{Fig.~#1\xspace}
\newcommand{\Sec}[1]{Sec.~#1\xspace}
\newcommand{\Tab}[1]{Tab.~#1\xspace}
\newcommand{\Eq}[1]{Eq.~#1\xspace}

\newcommand{\bx}{\boldsymbol{x}}
\newcommand{\bt}{\boldsymbol{\beta}}
\newcommand{\cD}{\mathcal{D}}
\newcommand{\cT}{\mathcal{T}}
\newcommand{\cO}{\mathcal{O}}
\newcommand{\cE}{\mathcal{E}}
\newcommand{\cM}{\mathcal{M}}
\newcommand{\mR}{\mathbb{R}}
\newcommand{\mP}{\mathbb{P}}

\renewcommand{\baselinestretch}{0.93}

\begin{document}

\title{\paperTitle}

\newcommand\Mark[1]{\textsuperscript#1}

\author{
	
	\IEEEauthorblockN{
		\textbf{Robert Falkenberg\Mark{1}, Benjamin Sliwa\Mark{1}, Nico Piatkowski\Mark{2} and Christian Wietfeld\Mark{1}}
	}

	\IEEEauthorblockA{
		\Mark{1}Communication Networks Institute, 
		\Mark{2}Department of Computer Science, AI Group\\ 
		TU Dortmund University, 44227 Dortmund, Germany\\
		e-mail:  $\{$Robert.Falkenberg, Benjamin.Sliwa, Nico Piatkowski, Christian.Wietfeld$\}$@tu-dortmund.de 			
	}
}

\maketitle

\begin{abstract}

Energy-aware system design is an important optimization task for static and mobile Internet of Things (IoT)-based sensor nodes, especially for highly resource-constrained vehicles such as mobile robotic systems.
For 4G/5G-based cellular communication systems, the effective transmission power of uplink data transmissions is of crucial importance for the overall system power consumption. Unfortunately, this information is usually hidden within off-the-shelf modems and mobile handsets and can therefore not be exploited for enabling green communication. Moreover, the dynamic transmission power control behavior of the mobile device is not even explicitly modeled in most of the established simulation frameworks.  
In this paper, we present a novel machine learning-based approach for forecasting the resulting uplink transmission power used for data transmissions based on the available passive network quality indicators and application-level information. The model is derived from comprehensive field measurements of drive tests performed in a public cellular network and can be parameterized for integrating all measurements a given target platform is able to provide into the prediction process.
In a comparison of three different machine learning methods, Random-Forest models thoroughly performed best with a mean average error of 3.166\,dB.
As the absolute sum of errors converges towards zero and falls below 1\,dB after 28 predictions in average, the approach is well-suited for long-term power estimations.

\end{abstract}

\IEEEpeerreviewmaketitle

\acresetall

\PrintCopyrightOverlay

\section{Introduction}

\begin{figure}[b!]  	
	\vspace{-10pt}
	\centering		  
	\subimport{fig/}{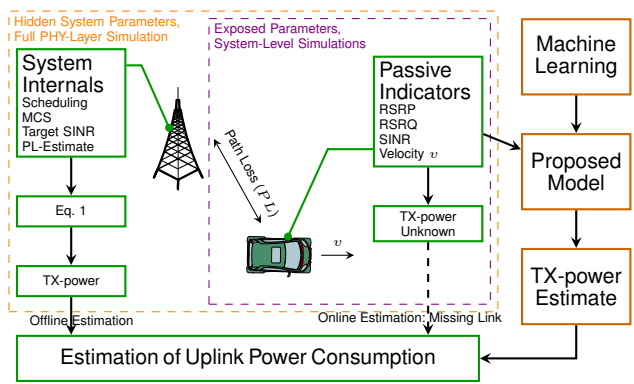}
	\caption{Uplink power control of \UE and its underlying system parameters are typically hidden to the application layers. However, this knowledge is crucial for predictions and estimations of the involved power consumption in energy-aware applications or system-level simulations, respectively. The proposed model derives this information from passive connectivity indicators.}
	\label{fig:scenario}
\end{figure}

Energy-constraints are a system-immanent challenge for most \IoT-based devices that directly affect how long a device can autonomously operate in the field. Therefore, green communication networks aim to optimize the energy-efficiency of data transmissions, since large amounts of the available power resources are spent on communication.
In cellular communication systems, such as \LTE and upcoming 5G networks, the consumed energy for data transmissions is mainly determined by the uplink transmission power~\cite{Miao2013}.
However, this crucial information is often not accessible for the application layer and can therefore not be leveraged for designing cross-layer and energy-aware communication mechanisms.
Moreover, even most established \LTE simulators do not consider detailed models of the consumed energy and focus on simple linear system-level considerations.
Although different models for estimating the power consumption for \LTE-based data transmissions exist, they rely on indicators that are not accessible %
in real-world scenarios as they are usually only used internally by the modem firmware and are hidden for users and developers (e.g. the number of assigned \RBs and carrier-specific constants).
In this paper, we present a novel data-driven model for predicting the \UL \txpower of \UE based on machine learning and empirical data from real traces.
The proposed prediction model is highly parameterizable and can therefore be configured to only integrate the parameters that are available on a given target platform (e.g. a specific \UE or a specific network simulator). Moreover, it is able to implicitly consider hidden variables that are correlated with the available information. Therefore, it is capable of filling the gap between full-featured physical layer measurements and system-level considerations. \Fig{\ref{fig:scenario}} provides an illustration of the prediction's data-flow.

This work presents the model in three variants, which differ in the number of available input features for the prediction.
\begin{itemize}
	\item The \emph{\fullModel} leverages the full feature set and exploits low-level information that is usually only available if the modem is interfaced directly using vendor-specific control commands (e.g. information about the neighboring cells).
	\item The \emph{\practicalModel} integrates indicators which are typically accessible on off-the-shelf \UEs using operating system \API abstractions. Therefore, it can provide additional information about the power consumption e.g. for live evaluation of resource-efficient data transmission schemes \cite{SliwaLiebigFalkenbergEtAl2018}, \cite{Sliwa/etal/2018a}.
	\item With the \emph{\simulationModel}, we provide a lightweight mechanism for simulation frameworks that allows to make a statement about the transmission power of the mobile device even when the simulator itself does not explicitly model \txpower control. In particular, this applies to many established system-level network simulators such as SimuLTE \cite{VirdisSteaNardini2015} and LTE-Sim \cite{PiroGriecoBoggiaEtAl2011}.
\end{itemize}

The paper is structured as follows: After giving an overview about different state-of-the-art power consumption models, we present the system model of our approach and relate it to the \acf{CoPoMo}~\cite{DuszaIdeChengEtAl2013}. Afterward, the setup of the empirical evaluations as well as the machine learning analysis is described and evaluated. Finally, we provide the underlying raw data and the capturing software in an Open Source way to guarantee a high level of transparency and enable the reader to synthesize customized models with respect to available indicators.

\section{Related Work}

Measuring and optimizing the energy consumption of \UEs is a major research topic as the consumed energy for data transmissions is strongly related to the battery lifetime of the mobile device.
Consequently, a wide range of different models has been proposed for analyzing the \UL and \DL power consumption of mobile devices.
In~\cite{HuangQianGerberEtAl2012}, the authors propose a power model for data transfer that assumes a linear dependency of power consumption and data rate. %
However, it lacks a consideration of the radio transceiver's power control.
The model in~\cite{JensenLauridsenMogensenEtAl2012} aggregates the power consumption of distinct components of \LTE \UE, such as base-band processing for transmission and reception and the connected radio transceiver units.
It is obtained from empirical measurements of real \UEs in the laboratory and depends on parameters such as instant transmission power, received power and the associated data rate in each direction.
The \copomo~\cite{DuszaIdeChengEtAl2013}, however, adds numerous context and system parameters, \eg, environment, mobility, and user activity into the power estimation.
Hence, the model does not only reflect the device itself but extends the scope to intrinsic and extrinsic influences that affect the device's average power consumption.
Since the main power draw is issued by data transmissions rather than receptions, the model is dominated by influencing factors, which affect the \txpower and transmission time in the \UL.
In the basic variant, it groups distinct power states into a four-state Markov chain.
Transitions between the states are expressed as service rates and arrival rates, which also depend on the radio conditions.
However, the model strongly depends on the distribution of required \txpower, which can be obtained either by ray-tracing simulations or from empirical data, such as~\cite{Joshi2017}.
While \copomo is perfectly suited to provide power estimations in offline simulations, an online application is prevented in most cases by the lack of knowledge about the instant \txpower at higher protocol layers.

In contrast to \COTS equipment, \acp{SDR} such as srsLTE~\cite{srsLTE2016} give full insight into the protocol stack and even allow hypothetical computations of values like \txpower at any time based on the complete knowledge of all involved parameters.
Advanced approaches even enable a prediction of resource assignments based on the momentary over-all cell load~\cite{Falkenberg2017c}.
However, freely available \SDR implementations often suffer from limited functionality, \eg, no support for handover, low transmission power, or imperfect channel estimation.
Hence they do not adequately represent typical \COTS equipment.

In some cases, however, the available system and context information on \COTS equipment may suffice to reconstruct or estimate hidden parameters from lower protocol layers by a machine-learning approach.
A general survey on the potential of machine learning for current and future mobile networks is given in~\cite{JiangZhangRenEtAl2017}. %
It is part of the related broader topic of anticipatory networking~\cite{Bui2017}, where network infrastructure and terminals act proactively according to learned or estimated user behavior and network characteristics.
While most of those approaches require cross-layer integration into multiple network components, this paper aims at a lightweight prediction model at application layer, which is suitable for both, online applications and system-level simulations, without any requirement for modifying existing network infrastructure.

\section{Machine-Learning Based Solution Approach}\label{sec:approach}

For performing a prediction of the \txpower, an identification of an indicator set is required in a first step, which reliably reflects the influencing factors on the power control of the UE.

These indicators have to be accessible at the application layer on most \LTE handsets for online applications and to be available in system-level simulators for offline applications.
However, taking the data rate as an example, such application-layer indicators might be blurred by a mixture of multiple influences on the radio system (\eg, signal quality and congestion) and the intermediate operating system (\eg, buffering and \TCP slow start mechanism).
Therefore, it requires a collection of multiple application-layer indicators which are influenced by the same underlying effect to deduce the true cause behind the curtain of abstraction.

The second step involves \ML to derive the complex relationship between the set of collected indicators, namely features, on the resulting \txpower of the device.
The required data for this task is obtained from real traces as explained in \Sec{\ref{sec:method}}.
Three different machine learning methods (\RandomF, \DeepL, and \RidgeR) of diverse power and complexity will be applied to rate the difficulty of the learning problem and identify the most lightweight solution for this task.
Furthermore, the model synthesis will be performed for different feature subsets, according to the typically available indicators in the addressed applications.
In this way, we also provide a trade-off between model complexity and accuracy, with respect to the demands and constraints that are given by the particular application scenarios.

\subsection{Relation of Downlink Indicators to Uplink Power Control}\label{sec:indicators}

According to the \LTE standard~\cite{36.213}, the \UE calculates its uplink transmission power $P_{tx}$ according to
\begin{align} \label{eq:analytic_p_tx}
P_{tx} = \min\left(\begin{array}{l}
P_{max},\\
P_{0} + 10 \log_{10}(M) + \alpha \cdot PL + \Delta_{MCS} + \delta
\end{array}
\right).
\end{align}
 First of all, it includes a compensation of the estimated path loss $PL$, which is weighted by a pre-configured Fractional Path Loss Compensation (FPC) factor $\alpha$. 
$P_0$ represents the requested \SINR per \RB for the reception by the \eNB.
Together with $\alpha$, these values are configured by the operator with respect to the environment and the building density.
Since the target power is always related to the \SINR per single \PRB, the total output power must be adjusted by the actual number of transmitting \RBs $M$ and the particularly increased \SINR of $\Delta_{MCS}$ for higher \acp{MCS}.
Finally, the \TPC formula includes a closed-loop component $\delta$.
It reflects cumulated instructions of the \eNB to the \UE to slightly increase or decrease its \txpower according to the actually received signal strength and signal quality at the \eNB.
The maximum output power of the \UE, however, is limited to $P_{max}$, which corresponds to  \SI{23}{\dBm} for class~3 \UE.

Although the latter is known to be a constant, the remaining parameters are either system- or context-dependent and are not accessible from \UE's application-layer.
However,  they either have direct or indirect influences on accessible indicators or at least can also be assumed or approximated as constants.
The path loss $PL$ may be tightly related to the \RSRP and \RSSI at the \UE.
Modulation and coding scheme $\Delta_{MCS}$ may additionally correlate to \RSRQ.
With this relationship and assuming the serving cell to not being congested, the number of assigned \RBs $M$ may correlate to the perceived \UL data rate at the \UE.
Furthermore, we assume $P_0$ to be constant for an operator, or at least for a homogeneous environment and expect $\delta$ to average to 0 for a properly configured cell.
The same assumption is applied on $\alpha$:
Due to its function, it is expected to be particularly constant for urban, suburban and rural environments.
It may be approximated by the currently used frequency band by the \UE and the number of visible neighbor cells.

\subsection{Machine Learning Methods and Model Synthesis}\label{sec:ML_models}
We consider three machine learning techniques to learn prediction models for the TX-power. The methods differ w.r.t. their complexity. On the one hand, complex models exhibit a high capacity which allows us to learn harder functions, resulting in a lower prediction error. This implies, however, that learning an actual parametric representation of the model is harder too, e.g., training a complex model requires more computational resources. On the other hand, computing the prediction with a complex model is computationally demanding and its parametric representation requires more memory. Let $\cD=\{(\bx^{(1)},y^{(1)}),(\bx^{(2)},y^{(2)}),\dots,(\bx^{(N)},y^{(N)})\}$ be a data set of size $N$. Here, $\bx\in\mR^d$ is a {\em feature vector} and $y\in\mR$ is the {\em label} that we want to predict, i.e., our goal is to estimate a function $f$ such that $f(\bx)\approx y$. 

\paragraph{Ridge Regression}
The ridge regression (RR) \cite{Hoerl/Kennard/2000a} is an $l_2$-regularized linear regression model with $f(\bx)=\langle \bt, \bx \rangle$. Learning RR models is done by minimizing the following regularized least-squares loss:
\begin{equation}
\min_{\bt\in\mR^d} \lambda \|\bt\|_2^2 + \underbrace{\frac{1}{N} \sum_{i=1}^N (y^{(i)}-f(\bx^{(i)}) )^2}_{\text{Mean squared error (MSE)}}.\label{eq:l2loss}
\end{equation}
The first term allows us to control the complexity---it penalizes models with overly large parameter values. To see this, observe that for $\lambda\to\infty$, the zero-vector will be optimal. In the other extreme ($\lambda=0$), the minimization problem is unregularized and we retrieve the model with minimal training error. By choosing a moderately small $\lambda$, we disallow the model to adapt itself to noise or other numerical particularities of the training data which results in better generalization performance---the model's prediction will be more accurate on unseen data \cite{Hastie/etal/2009a}. Linear models are particularly simple in that the number of parameters is equal to the number of features $d$. Due to strict convexity, the learning problem has a unique global minimizer. 

\paragraph{Random Forest}
Random forests (RF) \cite{Breiman/2001a} are a bootstrap \cite{Efron/1979a} of decision trees. Suppose that $T=(V,E)$ is a directed, tree structured graph. Each vertex $v\in V$ has (at most) two children $v_{\operatorname{left}}$ and $v_{\operatorname{right}}$. Moreover, for each vertex $v$, $\operatorname{val}(v)\in\mR$ is a real number, $\operatorname{idx}(v)\in\{1,2,\dots,m\}$ is a feature index, and $\operatorname{childs}(v)$ returns the number of children. The vertex function $f(v,\bx)$ is then
\begin{equation}\label{eq:decnode}
f(v,\bx) = \begin{cases}
f(v_{\operatorname{left}},\bx)&,\bx_{\operatorname{idx}(v)} \leq \operatorname{val}(v) \wedge \operatorname{childs}(v)=2\\
f(v_{\operatorname{right}},\bx)&,\bx_{\operatorname{idx}(v)} > \operatorname{val}(v) \wedge \operatorname{childs}(v)=2\\
\operatorname{val}(v)&,\operatorname{else}\;.
\end{cases}
\end{equation}
The prediction function of a single decision tree $T$ with root $v_0$ can then be written as $T(\bx)=f(v_0,\bx)$. Solving the corresponding learning problem is likely to have exponential runtime complexity. Thus, practical decision trees are grown heuristically by sequentially choosing $\operatorname{idx}(v)$ and $\operatorname{val}(v)$ such that $\sum_{i=1}^N (y^{(i)}-T(\bx^{(i)}))^2$ is minimized. Finally, a random forest is a set of trees $\cT=\{T_1,T_2,\dots\}$, where each tree $T_i$ is grown on a random bootstrap sample $\cD_i$ of the training data. The prediction of the forest is then the average ol all tree predictions: $f(\bx)=({1}/{M})\sum_{T\in\cT} T(\bx)$. Assuming that $D$ is the depth of the deepest tree, the forest has a worst-case storage complexity of $\cO(M 2^{D}-1)$.%

\paragraph{Deep Learning}
Deep learning (DL) \cite{Goodfellow/etal/2016a} is the machine learning technique which currently gets the highest attention from the research community and beyond. While classic (non-deep) approaches rely on hand-crafted features and various hyperparameters, deep learning methods aim at phrasing almost all parts of the model as differentiable function. Thus, numerical optimization methods can replace what was formerly done by hand. These methods work especially well in computer vision tasks, where a large number of semantically equivalent features is present, e.g., the pixel colors of an image. 
Compared to this setting, our number of features is rather small. However, we include this technique since it may discover unknown high-level features from our base features. More precisely, we use feed-forward neural networks with dropout \cite{Srivastava/etal/2014a}. Dropout is a regularization technique which randomly prohibits the update of some model parameters, thus, preventing the model from overfitting the training data. In contrast to linear models, the objective function is non-convex and the learning may get stuck in saddle points or weak local optima.

\section{Setup of the Empirical Evaluation}\label{sec:method}

This section describes the process of model generation, which includes the acquisition of training data from public networks and the subsequent machine learning procedure.

\subsection{Data Acquisition}

\begin{figure}[tbp]  	
	\vspace{\figureTopPadding}
	\centering		  
	\includegraphics[width=0.8\columnwidth]{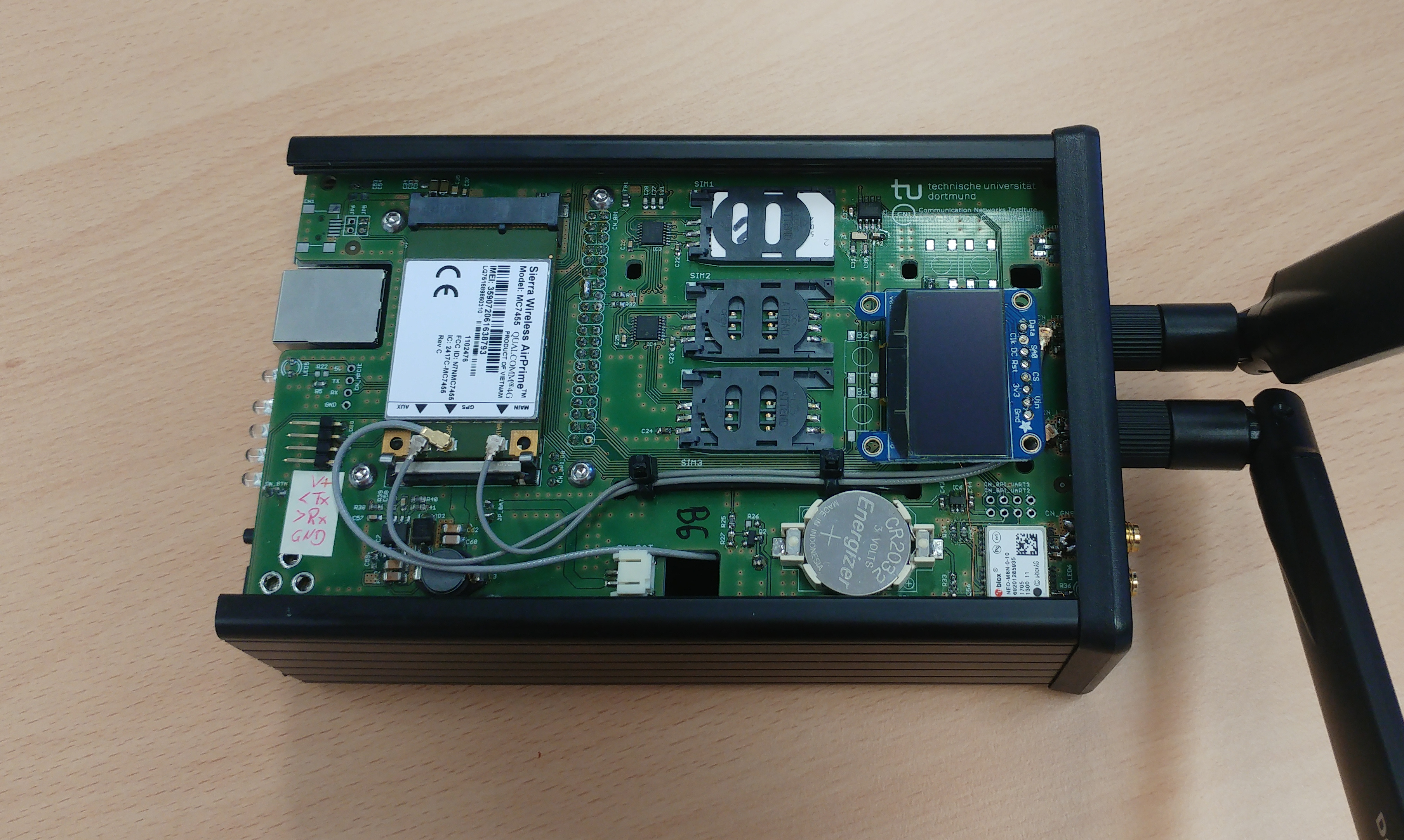}
	\caption{Photo of the embedded \ac{V2X} platform used for the measurements.}
	\label{fig:photo_measurement_box}
\end{figure}

In order to gather the necessary training data for model generation, mobile measurements of a public cellular network in Germany were performed.
They were taken by a developed embedded \ac{V2X} platform, shown in \Fig{\ref{fig:photo_measurement_box}}, which was mounted in the rear trunk of a car. %
This battery-powered system is based on an ARM-processor and includes a Sierra Wireless MC7455 \LTEA modem for network probing.
In contrast to \COTS smartphones, this modem exposes its momentary \txpower as well as further detailed information about its current state and the attached cellular network.
This includes a set of passive connectivity indicators, which are listed in  \Tab{\ref{tab:parameters}}.

\begin{table}[tbp]
	\centering
	\caption{Captured Features and Association to Application-Specific Prediction Models based on Full-Feature Set \modF, Practical~Sets \modPI/\modPII, and Simulation Set \modS.}
	\setlength\extrarowheight{2pt}	
	\begin{tabularx}{\columnwidth}{p{2.5cm}p{1.1cm}X}
		\toprule
		
		\textbf{Parameter} & \textbf{Model} & \textbf{Indicated Influences(s)} \\
		\midrule
		Velocity & \modF,\modPI,\modPII,\modS & Distortions by fast fading\\
		Upload size & \modF,\modPI,\modPII,\modS & Influence of TCP slow start\\
		\RSRP & \modF,\modPI,\modPII,\modS & Signal strength, distance\\
		\RSRQ, \SINR & \modF,\modPI,\modPII & Signal clarity, interference\\
		
		Datarate & \modF,\modPI & Signal strength, allocated \RBs $M$\\
		
		\RSSI & \modF & Signal strength, distance\\ 
		Frequency band & \modF & Environment~\cite{Joshi2017}\\
		
		Number of neighbor cells (intra/inter freq.) & \modF & Environment, cell density, interference\\
		
		Cell bandwidth & \modF & Exhaustion of \txpower headroom\\
		\bottomrule
	\end{tabularx}
	\label{tab:parameters}
	\vspace{\figureBottomPadding}	
\end{table}

The platform was instructed by our software (source code is available in~\cite{ZENODO}) to periodically upload a file via \HTTP to a web-server in intervals of \SI{30}{\second}.
During each transmission and in intervals of \SI{1}{\second}, the software sampled 31 parameters in total from the modem and the involved transmission process (\eg, data rate, file size).
In addition, the campaign was repeated with different upload file sizes of \SI{1}{\mega\byte}, \SI{3}{\mega\byte}, and \SI{5}{\mega\byte}.
The lower bound of the interval is motivated by the required query time for polling the \txpower together with the other parameters from the modem.
Although multiple values are queried in groups, single queries may take up to \SI{250}{\milli\second} depending on the modem's utilization.
Short transmissions, however, might be finished between two queries of \txpower and result in a useless sample due to no upload activity at query time.%

The upper bound, however, is motivated by the contract to conserve mobile traffic quota in favor of sampling over a large spatiotemporal domain but still performing measurements beyond the slow-start phase of a \TCP transmission.
We discuss the relevance of the slow start in \Sec{\ref{sec:results}}.

\begin{figure}[tbp]  	
	\vspace{\figureTopPadding}
	\centering		  
	\includegraphics[width=0.75\columnwidth]{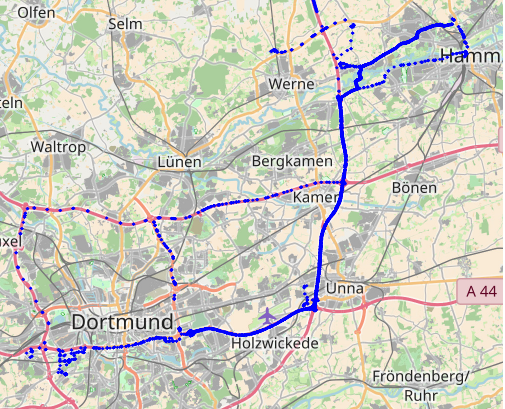}
	\caption{Road map with locations of all data samples of the measurement campaign between two larger cities in Germany. Each blue point represents an intermediate status logging of all measured variables (\cf \Tab{\ref{tab:parameters}}) during ongoing \UL transmissions.  (Map: \textcopyright OpenStreetMap contributors, CC BY-SA)}
	\label{fig:map}
	\vspace{\figureBottomPadding}	
\end{figure}

The road map in \Fig{\ref{fig:map}} shows  distinct measurement points along a trajectory through urban, suburban and rural environments, and hence provides data for an environment-independent model.
The main route has a length of \SI{44}{\kilo\meter} and stands out as dense line of trace points.
However, the data set also includes differing routes to avoid an over-fitting to the main path.
In total, 6172 samples were collected during this campaign.

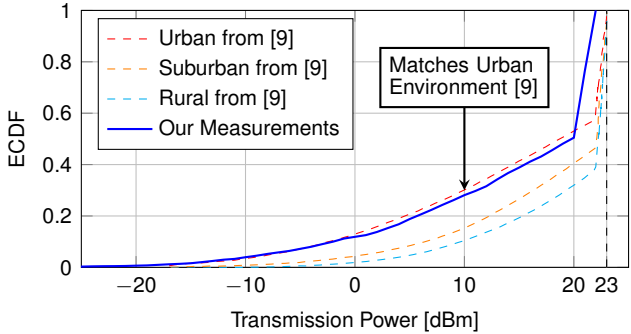
\begin{figure}[tb!]  	
	\vspace{\figureTopPadding}
	\centering		  
	\begin{tikzpicture}[
font=\sffamily\footnotesize,
]
\begin{sansmath}

\begin{axis}[
	width= 252.0pt,			%
	height=5cm,
	xmin=-25, xmax=25,
	ymin=0, ymax=1,
	scaled x ticks=false,		%
	extra x ticks={23},
	xlabel={Transmission Power [dBm]},
	ylabel={ECDF},
	ylabel near ticks,
	grid=both,
	legend cell align=left,
	legend pos=north west,
	]

	\tikzstyle{label} = [draw, fill=white]

	\addplot[color=red, dashed] table[col sep=comma, x = UrbanPower, y = UrbanPowerECDF] {\datapath/Joshi2017.csv};
	\addlegendentry{Urban from~\cite{Joshi2017}}

	\addplot[color=orange, dashed] table[col sep=comma, x = SuburbanPower, y = SuburbanPowerECDF] {\datapath/Joshi2017.csv};
	\addlegendentry{Suburban from~\cite{Joshi2017}}
	
	\addplot[color=cyan, dashed] table[col sep=comma, x = RuralPower, y = RuralPowerECDF] {\datapath/Joshi2017.csv};
	\addlegendentry{Rural from~\cite{Joshi2017}}

	\addplot[color=blue, thick] table[col sep=comma, x = txpower, y = ecdf] {\datapath/Dortmund_Hamm_TX_Power_ECDF.csv};
	\addlegendentry{Our Measurements}

	\draw[-,dashed] (axis cs:23,0) to (axis cs:23,1);

	\draw[stealth-, thick] (axis cs: 10, 0.3) to node[label, text width=2cm, at end] {Matches Urban Environment \cite{Joshi2017}} +(0cm,1.5cm);

\end{axis}

\end{sansmath}
\end{tikzpicture}
	\caption{\ECDF of the transmission power from our measurement campaign (blue) in Germany. For comparison, the dashed lines show the results from \cite{Joshi2017}, which were obtained from a mobile network operator in Sweden.}
	\label{fig:txpower-ecdf}
	\vspace{\figureBottomPadding}	
\end{figure}

In order to validate that the captured \txpower levels are representative for the environment, we plotted the \ECDF of our measurements into \Fig{\ref{fig:txpower-ecdf}}.
For comparison, we also included three distributions from~\cite{Joshi2017}, which reflect the \txpower distributions of different environments in Sweden.
That data was obtained from a large set of base stations with the support of a mobile network operator~\cite{Joshi2017}.
Our measurements match the distribution for urban environments from the reference set, although our the trajectory also includes suburban and rural areas.
However, the leading network operators in Germany typically expand their networks along motor highways, which explains the similarities to urban environments.

Note that the fork at \SI{0}{\dBm} is not issued by the environment but by post-processing the data:
Since the modem also reports a \txpower of \SI{0}{\dBm} in case of no activity, the dataset was filtered for those values to avoid ambiguities.
Furthermore, the modem's maximum \txpower was \SI{22}{\dBm}, which is \SI{1}{\dB} below the general limit of \SI{23}{\dBm}.
The power levels were validated in our laboratory by a Rohde\&Schwarz CMW500 radio communication tester.

\subsection{Application Scenarios and Involved Feature Subsets}
In order to provide tailored prediction models for online applications and offline simulations, we identified four different feature subsets \modF, \modPI, \modPII, \modS, which correspond to the particularly available indicators.
The \textbf{full-feature set} \modF includes all listed parameters from \Tab{\ref{tab:parameters}} and hereby covers all influences, that were outlined in \Sec{\ref{sec:indicators}}.
It serves as a reference to the subsequent subsets, since in practice the involved features are rarely available at application layer or require vendor-specific commands for access.

For \textbf{practical online-applications}, we defined feature subsets \modPI and \modPII, which have been labeled accordingly in \Tab{\ref{tab:parameters}}.
These are parameters, which are typically accessible on \COTS \UE, like smartphones or USB dongles.
Both sets only differ in the single parameter \emph{data rate}, which is only included in \modPII.
Although the data rate holds valuable implicit information about the number of allocated \RBs and consequently indicates the necessary additional \txpower, it is only available during an active transmission.
Therefore, a derived model, which involves this particular parameter, is only applicable for retrospective estimations of current and previous transmissions.
It is not suitable for predictions in idle mode.
This case, however, will be covered by the model, which is based on subset \modPI.

Since established simulators like SimuLTE and LTE-Sim provide even fewer indicators, a minimum \textbf{subset for simulations} \modS includes only three basic parameters \emph{velocity}, \emph{upload size}, and \emph{\RSRP}.

\subsection{Model Generation}\label{sec:ML_models}

For our experiments, we configured the models from Section~\ref{sec:ML_models} as follows:

Random forests are trained with $|\cT|=64$ trees, each having maximal depth 32. The model is saturated w.r.t. the number of trees, that is, we drove $|\cT|$ up to 128 where all forests with $|\cT|>64$ give essentially the same results.
Restricting the depth acts as a regularization term, hence preventing the model to memorize the complete training set and enhancing the generalization performance on previously unseen data.  %
Representing each of the $m$ nodes by $4$ values (left child, right child, threshold value, feature index), our forest can be represented by $d_{\operatorname{RF}}= m\times 4 = 1044992$ values. 

Ridge regression models are trained with $\lambda=10^{-3}$. We conducted a parameter grid search with $\lambda=10^{-i}$ for $i\in\{0,1,2,3,4,5\}$, where all choices resulted in a reasonable generalization performance. The number of model parameters is $d_{\operatorname{RR}}=12$ (one per feature plus one bias term).%

Our deep learning model consists of three fully connected hidden layers, each with 64 hidden nodes and non-linearity via rectified linear units (ReLU). We use an increasing dropout rate per layer which is motivated by various empirical and theoretical insights \cite{Gal/etal/2017a}. More precisely, $\mP_{\operatorname{Dropout}}=0.1\times i$ on layer $i\in\{1,2,3\}$. The parameters are estimated by stochastic gradient descent with an adaptive learning rate and standard hyperparameters ($\epsilon=10^{-8},\rho=0.99$). The total number of learnable parameters is $d_{\operatorname{DL}}=12 \times 64 + 64^2 + 64^2 + 64=9024$. 

All models are learned with RapidMiner 8.1 and the corresponding process files can be found online \cite{ZENODO}.

\section{Numerical Results}\label{sec:results}
We start by assessing the importance of all features w.r.t. the TX-power. To this end, we binned all values into 10 intervals $b_1,b_2,\dots,b_{10}$ of equal width and computed the mutual information (MI) between each feature and the TX-power, $\operatorname{MI}(X,Y)=\sum_{i=1}^{10} \sum_{j=1}^{10} p_{i,j}\log ({p_{i,j}}/{p_ip_j})$.
Here, $Y=\operatorname{TX-power}$, $X$ is any of the \modF features, $p_i=\hat{p}(X=b_i)$ is the relative number of data points in which feature $X$ is in bin $b_i$ and $p_{i,j}=\hat{p}(X=b_i,\operatorname{TX-power}=t_j)$ is the number of cases in which $X=b_i$ and $\operatorname{TX-power}=t_j$ simultaneously. The MI measure how dependent two variables are on each other. In contrast to plain correlation, MI is able to capture non-linear dependence. 

The MI values in descending order are \RSRP $0.589$, \RSSI $0.543$, num. of intra freq. neighbor cells $0.217$, \SINR $0.212$, \RSRQ $0.153$, data rate $0.090$, velocity $0.076$, upload size $0.041$, num. of inter freq. neighbor cells $0.013$, frequency band $0.008$, and cell bandwidth $0.007$. The most important parameters are \RSRP and \RSSI, which imply the signal path loss of the downlink signal and particularly provides an estimate of the path loss in the uplink direction.
However, these two indicators carry redundant information, since the models based on smaller feature sets without \RSSI still provide very accurate results.
The same applies to the indicator group of \RSRQ, \SINR, and number of neighbor cells (intra freq.), which imply the signal quality, however with a smaller importance for the full model.
Frequency band, cell bandwidth, and inter-frequency neighbors have only negligible relevance for the resulting model. 

We saw that the most important feature is contained in all feature seats under consideration. It is hence reasonable to expect that all feature subsets contain enough information for a decent predictive quality. 
Now, we provide results regarding the prediction error of our machine learning models, an analysis of side effects, and the cumulation of errors. All results are 10-fold cross-validated \cite{Hastie/etal/2009a}, i.e., our data set is partitioned into 10 sets. Each model is trained 10 times, where in each run, the model is trained on 9 sets and tested on the remaining set. The prediction errors of these 10 runs are then averaged. Cross-validated results are more reliable than results gathered from a single train/test split, since a single split might be strongly in favor or strongly to the disadvantage of a method. By averaging over multiple runs, such artifacts are avoided. 

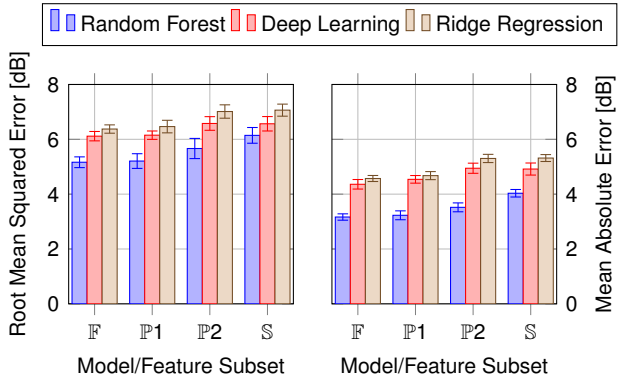
\begin{figure}[tb!]  	
	\vspace{\figureTopPadding}
	\centering		  
	\begin{tikzpicture}[
	font=\sffamily\footnotesize,
	]
\begin{sansmath}

\pgfplotstableread[row sep=\\,col sep=&]{
	SUBSET & RF & RFE & DL & DLE & RR & RRE \\
	\modF   & 5.164 & 0.196 & 6.115 & 0.171 & 6.375 & 0.152 \\
	\modPI  & 5.207 & 0.268 & 6.151 & 0.153 & 6.465 & 0.229 \\
	\modPII  & 5.665 & 0.368 & 6.578 & 0.249 & 7.014 & 0.244 \\
	\modS	& 6.144 & 0.287 & 6.567 & 0.263 & 7.064 & 0.220 \\
}\RMSE

\pgfplotstableread[row sep=\\,col sep=&]{
	SUBSET & RF & RFE & DL & DLE & RR & RRE \\
	\modF   & 3.166 & 0.116 & 4.360 & 0.174 & 4.571 & 0.110 \\
	\modPI  & 3.229 & 0.162 & 4.540 & 0.139 & 4.676 & 0.147 \\
	\modPII  & 3.519 & 0.162 & 4.945 & 0.185 & 5.302 & 0.150 \\
	\modS	& 4.033 & 0.137 & 4.916 & 0.220 & 5.317 & 0.126 \\
}\MAE

    \begin{axis}[
		xshift=3.5cm,
		ybar=0pt,
		width= 130.0pt,			%
		height=4.5cm,
		enlarge x limits=0.15,
		ymin=0, ymax=8,
		symbolic x coords={\modF,\modPI,\modPII, \modS},
		bar width=0.2cm,
		xtick=data,
		xlabel={Model/Feature Subset},
		ylabel={Mean Absolute Error [dB]},
		ylabel near ticks,
		yticklabel pos=right,
		grid=both,
		legend cell align=left,
		legend pos=north west,
		legend style={legend columns=-1,anchor=north west},	%
		]
		\addplot+[] plot[error bars/.cd, y dir=both, y explicit] table[x=SUBSET,y=RF, y error=RFE]{\MAE};
		\addplot+[] plot[error bars/.cd, y dir=both, y explicit] table[x=SUBSET,y=DL,y error=DLE]{\MAE};
		\addplot+[] plot[error bars/.cd, y dir=both, y explicit] table[x=SUBSET,y=RR,y error=RRE]{\MAE};
\end{axis}

    \begin{axis}[
		ybar=0pt,
		width= 130.0pt,			%
		height=4.5cm,
		enlarge x limits=0.15,
		ymin=0, ymax=8,
	    symbolic x coords={\modF,\modPI,\modPII, \modS},
	    bar width=0.2cm,
	    xtick=data,
		xlabel={Model/Feature Subset},
		ylabel={Root Mean Squared Error [dB]},
		grid=both,
		legend cell align=left,
		legend style={legend columns=-1,anchor=north west,at={(-0.7cm,4cm)},fill=white},	%
	    ]
	    \addplot+[] plot[error bars/.cd, y dir=both, y explicit] table[x=SUBSET,y=RF, y error=RFE]{\RMSE};
	    \addplot+[] plot[error bars/.cd, y dir=both, y explicit] table[x=SUBSET,y=DL,y error=DLE]{\RMSE};
	    \addplot+[] plot[error bars/.cd, y dir=both, y explicit] table[x=SUBSET,y=RR,y error=RRE]{\RMSE};
	    \legend{Random Forest, Deep Learning, Ridge Regression}
    \end{axis}

\end{sansmath}
\end{tikzpicture}
	\caption{Cross-validated error of trained prediction models for each feature subset (\modF, \modPI, \modPII, \modS) and each machine learning method (Random Forest, Deep Learning, Ridge Regression) in terms of \RMSE (left) and \MAE (right). Lower is better. %
}
	\label{fig:accuracy}
	\vspace{\figureBottomPadding}	
\end{figure}

Results in terms of prediction error (in [\si{\dB}]) of the trained models are provided in \Fig{\ref{fig:accuracy}}. 
The left plot shows the \RMSE (the square root of the MSE from equation~\ref{eq:l2loss}) for all three models $f\in\{\operatorname{DL},\operatorname{RF},\operatorname{RR}\}$.
On the right, we present the \MAE (right side) which is defined as $(1/N)\sum_{i=1}^N |y^{(i)}-f(\bx^{(i)})|$. 
The RMSE weights large prediction errors heavier and tiny prediction errors slightly, while the MAE treats all errors equally. 
Error bars are displaying the standard deviation computed over the  10 cross-validation runs.

In any case, the deviations are below \SI{0.4}{\dB} with maximum values of \SI{0.368}{\dB} and \SI{0.220}{\dB} for \RMSE and \MAE, respectively.
This indicates a good model fit to unknown and independent data, which are not included in the training set.
According to the results, \RandomF performs best with an \MAE of \SI{3.166}{\dB} in the full feature subset \modF.
However, even in the most compact feature subset for simulations \modS, the error raises only moderately by less than \SI{1}{\dB} to \SI{4.033}{\dB}.
Therefore, \modS-models have only a slightly increased error compared to the other models, but rely only on the three input features velocity, upload size, and \RSRP.
This makes it a valuable option for system-level simulations, which model mobile networks with a very coarse-grained amount of detail.

Comparing the different \ML methods, \RandomF outperforms the other methods on all feature subsets.
However, all results have the same order of magnitude, which suggest that the residual errors are not an artifact of the utilized \ML method, but a consequence of the uncertainty induced by an incomplete view of all influencing factors.
On the other hand, as \RidgeR produces the largest errors in this setup, a non-linear relationship of input features and resulting \txpower is suggested.

\begin{figure}[tb!]  	
	\vspace{\figureTopPadding}
	\centering		  
	\begin{tikzpicture}[
font=\sffamily\footnotesize,
]
\begin{sansmath}

\begin{axis}[
	width= 252.0pt,			%
	height=6cm,
	xmin=-130, xmax=-65,
	ymin=-15, ymax=25,
	scaled x ticks=false,		%
	extra y ticks={23},
	xlabel={RSRP [dBm]},
	ylabel={Transmission Power [dBm]},
	ylabel near ticks,
	grid=both,
	legend cell align=left,
	legend pos=south west,
	]

	\tikzstyle{label} = [draw, fill=white]

	\addplot[color=blue] plot[error bars/.cd, y dir=both, y explicit] table[col sep=comma, x = X, y = Y, y error = CIpos] {\datapath/Dortmund_Hamm_RSRP_TX_Power_Averaged_DH1.csv};
\addlegendentry{1 MB Upload}
	\addplot[color=red] plot[error bars/.cd, y dir=both, y explicit] table[col sep=comma, x = X, y = Y, y error = CIpos]  {\datapath/Dortmund_Hamm_RSRP_TX_Power_Averaged_DH3.csv};
\addlegendentry{3 MB Upload}
	\addplot[color=green] plot[error bars/.cd, y dir=both, y explicit] table[col sep=comma, x = X, y = Y, y error = CIpos]  {\datapath/Dortmund_Hamm_RSRP_TX_Power_Averaged_DH5.csv};
	\addlegendentry{5 MB Upload}

	\draw[-,dashed] (axis cs:-130,23) to (axis cs:-65,23);
	\draw[-,dashed] (axis cs:-100,-15) to (axis cs:-100,24);
	
	\node[label, thick, text width=2.8cm, align=center] at (axis cs: -115,15) {TX-power saturated,\\reduced data rate};
	\node[label, thick, text width=3.5cm, rotate=-45] at (axis cs: -85,-2) {TCP slow start scatters\\TX-power w.r.t. upload size};

\end{axis}

\end{sansmath}
\end{tikzpicture}
	\vspace*{-5pt}
	\caption{Relationship between \RSRP and \txpower grouped by different upload sizes. Larger uploads lead to higher \txpower levels, as the \TCP slow start algorithm reaches higher data rates and involves the transmission of more \RBs in parallel. The effect decreases as \txpower gets saturated.}
	\label{fig:rsrp_prediction}
	\vspace{\figureBottomPadding}	
\end{figure}
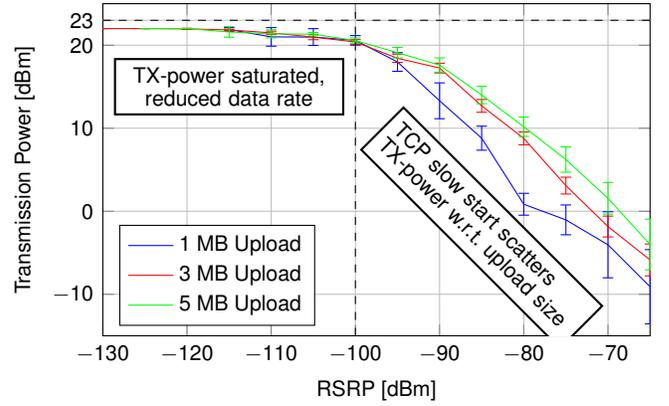

This is also visible in \Fig{\ref{fig:rsrp_prediction}}, which depicts the (single-feature) relationship between \RSRP and resulting \txpower for different upload sizes.
The figure shows the averaged \txpower for bins of \RSRP values with a width of \SI{5}{\dB}.
The bins $b_r$ are aligned to \RSRP values $r$, where $(r\mod 5) = 0$ and include values in the interval $[r, r+5)$.
Error bars are showing the $0.95$ \CIs of the observed samples.
Two areas are visible, which differ in the extent of \CI and the impact of data size on the average \txpower.
They can be coarsely separated by an \RSRP value of \SI{-100}{\dBm}.
Samples below that threshold, which generally indicate low or poor radio conditions, show very tight \CIs and congruent \txpower levels.
For larger \RSRP values, however, the \txpower is affected by an increased degree of scattering and the mean value becomes data-size dependent.
As pointed out in \Sec{\ref{sec:approach}} and \Eq{\ref{eq:analytic_p_tx}}, the \txpower depends on the number of assigned \RBs $M$.
In spite of its direct relationship to the throughput, the latter is also affected by $\Delta_{MCS}$ and buffering by the operating system.
Therefore, the data rate is only reflecting the average resource utilization on physical layer but not the instant situation at subframe level.
Given a fixed $\Delta_{MCS}$ in \Eq{\ref{eq:analytic_p_tx}}, the uncertainty of $M = 1\ldots 100$ leads to a possible scattering of $P_{tx}$ in the range of \SI{20}{\dB} for a \SI{20}{\mega\hertz} cell.

A possible explanation for the dependency of the data size is the \TCP slow start mechanism.
In this case, the transmitter slowly increases its data rate until the maximum data rate is reached.
By the same token, the amount of assigned \RBs $M$ also slowly increases as long as spare resources are available.
Consequently, small payloads, \eg, \SI{1}{\mega\byte}, rarely reach full \RB utilization, because the transfer is already finished in the slow start region.
As the data size increases, the average throughput also grows towards a full \RB allocation, and pushes the average transmission power to a higher level.
However, in case of poor radio conditions (left side of \Fig{\ref{fig:rsrp_prediction}}), the number of \RBs is quickly saturated by the maximum \txpower of the \UE, which limits the maximum data rate to a much smaller value.
Therefore, the slow start quickly reaches the maximum data rate even in case of small payloads.

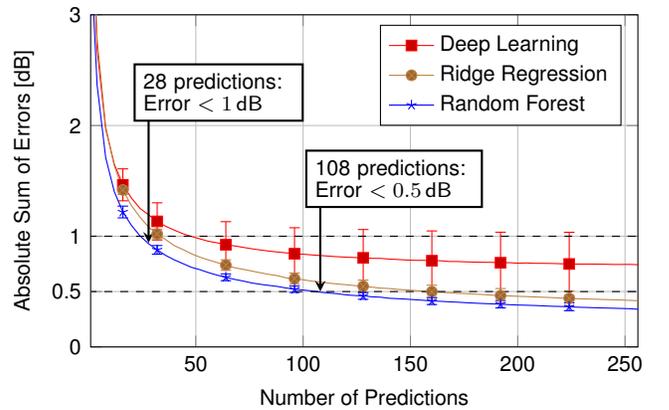
\begin{figure}[b!]  	
	\vspace{-10pt}
	\centering		  
	\begin{tikzpicture}[
font=\sffamily\footnotesize,
]
\begin{sansmath}

\begin{axis}[
	width= 252.0pt,			%
	height=6cm,
	xmin=1, xmax=256,
	ymin=0, ymax=3,
	scaled x ticks=false,		%
	extra y ticks={0.5},
	xlabel={Number of Predictions},
	ylabel={Absolute Sum of Errors [dB]},
	ylabel near ticks,
	grid=both,
	legend cell align=left,
	legend pos=north east,
	]

	\tikzstyle{label} = [draw, fill=white]

	\pgfplotsset{cycle list shift=1}
	\addplot[color=red,no markers,forget plot] plot table[col sep=comma, x = terms, y = NN, y error = NN_SDEV] {\datapath/err_sum.csv};

	\addplot[color=brown,no markers,forget plot] plot table[col sep=comma, x = terms, y = LIN, y error = LIN_SDEV]  {\datapath/err_sum.csv};

	\addplot[color=blue,no markers,forget plot] plot table[col sep=comma, x = terms, y = RF, y error = RF_SDEV]  {\datapath/err_sum.csv};
	\addplot+[color=red, restrict expr to domain={
		x==16||
		x==32||
		x==64||
		x==96||
		x==128||
		x==160||
		x==192||
		x==224
		}{1:1}] plot[error bars/.cd, y dir=both, y explicit] table[col sep=comma, x = terms, y = NN, y error = NN_SDEV] {\datapath/err_sum.csv};
	\addlegendentry{Deep Learning}
	
	\addplot+[color=brown,restrict expr to domain={
	x==16||
	x==32||
	x==64||
	x==96||
	x==128||
	x==160||
	x==192||
	x==224
}{1:1}] plot[error bars/.cd, y dir=both, y explicit] table[col sep=comma, x = terms, y = LIN, y error = LIN_SDEV] {\datapath/err_sum.csv};
\addlegendentry{Ridge Regression}

	\addplot+[color=blue,restrict expr to domain={
	x==16||
	x==32||
	x==64||
	x==96||
	x==128||
	x==160||
	x==192||
	x==224
}{1:1}] plot[error bars/.cd, y dir=both, y explicit] table[col sep=comma, x = terms, y = RF, y error = RF_SDEV] {\datapath/err_sum.csv};
	\addlegendentry{Random Forest}

	\draw[-,dashed] (axis cs:0,1) to (axis cs:256,1);
	\draw[-,dashed] (axis cs:0,0.5) to (axis cs:256,0.5);
	
    \draw[stealth-,thick] (axis cs: 28,0.94) to node[label, at end, anchor=west, xshift=-0.2cm, text width=2cm] {28 predictions:\\Error $<\SI{1}{\dB}$} +(0cm,2cm);
    \draw[stealth-,thick] (axis cs: 108,0.5) to node[label, at end, anchor=west, xshift=-0.2cm, text width=2.1cm] {108 predictions:\\Error $<\SI{0.5}{\dB}$} +(0cm,1.5cm);

\end{axis}

\end{sansmath}
\end{tikzpicture}
	\vspace*{-5pt}
	\caption{Cumulated absolute error of the proposed prediction models in relation to the number of executed predictions. In average the Random-Forest model undershoots the error mark of \SI{1}{\dB} after 28 predictions. Therefore, in long-term applications over- and underestimations eliminate each other and provide an accurate estimate for the average \txpower. }
	\label{fig:err_sum}
\end{figure}

Besides the accuracy of single predictions, the cumulative behavior of the prediction error has a large relevance for long-term applications. 
Therefore, we designed a numerical simulation to estimate the absolute sum of errors in relation to the number of executed predictions \Fig{\ref{fig:err_sum}}. 
The points on the curve of each method are generated as follows: For each combination of cross-validation run and ``Number of predictions'' $l$ (the abscissae), we draw $l$ random indices $i_1,i_2,\dots,i_l$ and compute the cumulative error $e_j^l=(1/l)\sum_{i=1}^l y_{i_l} - \hat{y}_{i_l}$, where $y$ is the true value and $\hat{y}$ the prediction. This is repeated $\cM=10^4$ times and averaged $\cE_l=(1/\cM)\sum_{i=1}^\cM |e_j^l|$ where $|\cdot|$ denotes the absolute value. Taking the absolute value is required to prevent that errors from the $\cM$ independent runs cancel each other out. Thus, the value $\cE_l$ is an estimate to the expected absolute error that we see, if we cumulate the pointwise errors of $l$ predictions. It is necessary to average the result of this computation on each test set of the cross-validation to avoid overly pessimistic or overly confident estimators, i.e., the final ordinate for the abscissa $l$ is $E^*_l=(1/10)\sum_{k=1}^{10} E^k_l$ where $E^k_l$ is computed on the $k$-th cross-validation run. The estimation is carried out for each method and each $l\in\{1,2,4,8,12,\dots,256\}$ to yield the curves shown in \Fig{\ref{fig:err_sum}}. The same procedure is used to estimate the standard deviation of the cumulated errors of $l$ predictions, which are shown in the plot via error bars.

With Random Forest as an example, the cumulated absolute error undershoots in average after 28 predictions an error of \SI{1}{\dB}, after 108 predictions it even falls below \SI{0.5}{\dB}.
As the deviation is very small, the proposed approach is suitable for long-term applications, since the cumulated prediction error does not diverge over time.

\section{Conclusion}

In order to rate the power consumption of \UEs during uplink transmissions in mobile networks,
knowledge about the instant \txpower is required.
As this information is not accessible on most mobile devices and in many system-level simulators, this paper closes the gap by a customizable machine-learning approach.
The \txpower is estimated on the basis of passive indicators, such as \RSRP, velocity, and data rate, which are generally accessible in the mentioned applications.
The required trace data is obtained from excessive drive tests that include periodic uplink transmissions over a public mobile network.
Three machine learning methods (Random Forrest, Deep Learning, and Ridge Regression) were applied on four different feature subsets of the obtained data to provide the most accurate and lightweight predictor according to the amount of available indicators on the target platform.
The full-featured model provides predictions with an \MAE of \SI{3.166}{\dB} by using Random Forests.
However, even when limiting the input features to velocity, upload size, and \RSRP, the \MAE only raises to \SI{4.033}{\dB}.
For long-term applications the model maintains stability, since over- and underestimations eliminate each other and the cumulated error sum converges towards 0.

In future work, we will further investigate the impact of the \TCP congestion control mechanism and include measurements of unbuffered \UDP transmissions into the dataset.
Furthermore, we will integrate the prediction models into \copomo for online estimations of the momentary power consumption.

\section*{Acknowledgment}

Part of the work on this paper has been supported by Deutsche Forschungsgemeinschaft (DFG) within the Collaborative Research Center SFB~876 ``Providing Information by Resource-Constrained Analysis'', projects A1, A4, and B4.

\renewcommand{\baselinestretch}{0.945}

\bibliographystyle{IEEEtran}
\bibliography{Manuscript}

\end{document}